\documentclass[aps,twocolumn,pra]{revtex4-1}
\usepackage{graphicx}
\usepackage{amsmath}

\begin{document}

\title{Super-Klein tunneling of Klein-Gordon particles}

\author{Kihong Kim}
\email{khkim@ajou.ac.kr}
\affiliation{Department of Energy Systems Research and Department of Physics, Ajou University, Suwon 16499, Korea}
\affiliation{School of Physics, Korea Institute for Advanced Study, Seoul 02455, Korea}

\begin{abstract}
We study the total transmission of quantum particles
satisfying the Klein-Gordon equation through a potential barrier based on the classical wave propagation theory.
We deduce an analytical expression for the wave impedance for Klein-Gordon particles. From the condition that
the impedance is matched throughout the space, we show that
super-Klein tunneling, which refers to the omnidirectional total transmission of particles through a potential barrier
when the energy of the particle is equal to the half of the barrier potential, should occur in
these systems.
We also derive a condition for total transmission of Klein-Gordon particles in the presence of both scalar and vector potentials
and discuss the influence of weak scattering at the interfaces on super-Klein tunneling.
The theoretical predictions are confirmed by explicit numerical calculations of the transmittance based on the invariant imbedding method.
\end{abstract}

\maketitle

\section{Introduction}
Among many interesting properties of quantum-mechanical particles satisfying the Dirac equation, Klein tunneling, which refers to
the phenomenon that Dirac fermions entering a large potential barrier normally
are transmitted almost completely as the barrier becomes higher and wider, has been the focus of much recent research, especially
in the context of massless Dirac electrons in graphene \cite{klein,kats,shg,young}. This has stimulated interest in various total transmission phenomena in the systems of
pseudo-relativistic particles in solid-state materials and in analogous classical wave systems \cite{jalil,bai,nicol,nguyen,yang,oca2,yang2}.
Super-Klein tunneling refers to the omnidirectional total transmission of quantum particles through a potential barrier
when the energy of the incident particle is precisely equal to the half of the barrier potential.
This phenomenon was found to exist in the systems satisfying the pseudospin-1 Dirac equation, but not to appear in those
satisfying the isotropic pseudospin-1/2 Dirac equation such as graphene \cite{shen,urb,fang,oca}.
Recently, it has been demonstrated that super-Klein tunneling is also possible in anisotropic pseudospin-1/2 Dirac materials \cite{jpc}.
In this paper, we demonstrate for the first time that super-Klein tunneling occurs in the systems satisfying the Klein-Gordon equation,
which is generally believed to describe spin-0 bosons,
regardless of the mass of the particles.
Starting from the Klein-Gordon equation, we deduce an analytical expression for the wave impedance and show that
the condition for super-Klein tunneling is nothing but the condition that the impedance is matched throughout the system.
We also derive a condition for total transmission of Klein-Gordon particles in the presence of both scalar and vector potentials
and discuss the influence of weak scattering at the interfaces on super-Klein tunneling.

\section{Theory}
The Klein-Gordon equation is obtained by quantizing the relativistic energy-momentum relation
$E^2=p^2c^2+m^2c^4$.
In the presence of one-dimensional scalar potential $U=U(x)$ and vector potential ${\bf A}=A_y(x) {\hat {\bf y}}$,
which is related to an external magnetic field perpendicular to the $xy$ plane, ${\bf B}=B_0{\hat {\bf z}}$,
by $B_0(x)=dA_y(x)/dx$, the time-independent Klein-Gordon
equation for charged particles moving in the $xy$ plane takes the form
\begin{eqnarray}
\left[\left(E-U\right)^2-m^2c^4\right]\psi=c^2{\pi_x}^2\psi+c^2{\pi_y}^2\psi,
\label{eq:k1}
\end{eqnarray}
where the components of the kinetic momentum operator, $\pi_x$ and $\pi_y$, are
\begin{eqnarray}
\pi_x=\frac{\hbar}{i}\frac{d}{dx},~~\pi_y=\hbar k_y-\frac{q}{c}A_y.
\label{eq:mom}
\end{eqnarray}
$k_y$ is the $y$ component of the wave vector and $q$ ($=-e$) is the particle charge.
By substituting Eq.~(\ref{eq:mom}) into Eq.~(\ref{eq:k1}), we obtain
\begin{eqnarray}
\frac{d^2\psi}{dx^2}+\left[\frac{\left(E-U\right)^2-m^2c^4}{\hbar^2c^2}-\left({k_y}
+\frac{eA_y}{\hbar c}\right)^2\right]\psi=0.
\label{eq:we1}
\end{eqnarray}
We assume that a plane wave described by the wave function
$\psi$ is incident obliquely at an angle $\theta$ from the free region $x>L$ where $U=A_y=0$
onto the region in $0\le x\le L$ where $U=U(x)$ and $A_y=A_y(x)$,
and then transmitted to the free region $x<0$ where $U=A_y=0$.
The wave number $k$ and the negative $x$ component of the wave vector, $k_x$, in the incident and transmitted regions
and the $y$ component of the wave vector, $k_y$, which is the constant of the motion, are given by
\begin{eqnarray}
k=\frac{\sqrt{E^2-m^2c^4}}{\hbar c},~k_x=k\cos\theta,~k_y=k\sin\theta,
\end{eqnarray}
where we have assumed that $\vert E\vert > m c^2$.
Then the Klein-Gordon equation takes the simple form
\begin{eqnarray}
\frac{d^2\psi}{dx^2}+{k_x}^2\eta^2\psi=0,
\end{eqnarray}
where the parameter $\eta$, which plays the role of the wave impedance, is given by
\begin{eqnarray}
\eta^2=\frac{\epsilon-\beta^2}{\cos^2\theta},~~\epsilon=1+\frac{U\left(U-2E\right)}{E^2-m^2c^4},~~\beta=\sin\theta+a.
\label{eq:imped1}
\end{eqnarray}
The dimensionless vector potential $a$ is defined by $a=eA_y/(\hbar ck)$.

We see easily that in the incident and transmitted regions, $\eta$ is identically equal to 1.
Therefore, if $\eta$ is unity in the region $0\le x\le L$ as well, the impedance is matched throughout the space
and there will be no wave reflection.
In order to have an omnidirectional transmission, we need to have $\eta$ independent of $\theta$.
This is possible only when $a=0$ and $E=U/2$ so that the $\cos^2\theta$ term in the denominator of Eq.~(\ref{eq:imped1})
can be cancelled out. In this case, $\eta$ is equal to 1 for all $\theta$,
which implies that the waves incident on the potential barrier in an arbitrary direction are transmitted completely
regardless of the barrier width and the particle mass.
A similar phenomenon has been found in the systems satisfying the pseudospin-1 Dirac equation and has been
termed super-Klein tunneling \cite{shen,urb,fang,oca}. To the best of our knowledge, this omnidirectional total transmission of Klein-Gordon particles
through a barrier has never been pointed out before.

In Ref.~\cite{gun}, it has been shown that when $U$ is greater than $E$, the system satisfying the Klein-Gordon equation behaves like
a negative refractive index medium. In particular, the case with $E=U/2$ corresponds to the medium with the refractive index $n$ equal to $-1$,
when the wave is incident from the free region with $n=1$. Then the concept of {\it complementary materials} can be invoked to explain
super-Klein tunneling in the present case, as has been done in \cite{fang} in the pseudospin-1 case. In addition,
Veselago focusing of particle flow should also occur, similarly to the pseudospin-1/2 snd pseudospin-1 cases \cite{shen,fang,alt}.

The ordinary Klein tunneling, which occurs for Dirac particles at $\theta=a=0$, does not occur in the Klein-Gordon case.
However, in the presence of the vector potential, total transmission can occur at a special incident angle $\theta_c$ satisfying
\begin{eqnarray}
\sin\theta_c=\frac{1}{2a}\frac{U\left(U-2E\right)}{E^2-m^2c^4}-\frac{a}{2},
\label{eq:ang}
\end{eqnarray}
which has been obtained by setting $\eta=1$ in Eq.~(\ref{eq:imped1}).
We note that $\theta_c$ can be tuned continuously by tuning the value of $E$, $U$ or $a$. The symmetry with respect to the sign
of $\theta$ is broken in this case.

We can solve the wave equation, Eq.~(\ref{eq:we1}), using
the invariant imbedding method \cite{kly,kim}. In this method, we first calculate the reflection and transmission coefficients $r$ and $t$ defined
by the wave functions in the incident and transmitted regions:
\begin{eqnarray}
\psi\left(x,L\right)=\left\{\begin{array}{ll}
  e^{-ik_x\left(x-L\right)}+r(L)e^{ik_x\left(x-L\right)}, & x>L \\
  t(L)e^{-ik_x x}, & x<0
  \end{array},\right.
\end{eqnarray}
where $r$ and $t$ are regarded as functions of $L$.
Following the procedure described in Ref.~\cite{kim} to derive the equations for $r$ and $t$, we obtain
\begin{eqnarray}
\frac{1}{k}\frac{dr}{dl}=-\frac{i\cos\theta}{2}\left(r-1\right)^2
+\frac{i}{2\cos\theta}\left(\epsilon-\beta^2\right)\left(r+1\right)^2,\nonumber\\
\frac{1}{k}\frac{dt}{dl}=-\frac{i\cos\theta}{2}\left(r-1\right)t
+\frac{i}{2\cos\theta}\left(\epsilon-\beta^2\right)\left(r+1\right)t.
\label{eq:imb1}
\end{eqnarray}
For any arbitrary functional forms of $U$ and $A_y$ and for any values of $kL$ and $\theta$,
we can integrate these equations from $l=0$ to $l=L$ using the initial conditions $r(0)=0$ and $t(0)=1$
and obtain $r(L)$ and $t(L)$. The reflectance $R$ and the transmittance $T$ are obtained using $R=\vert r\vert^2$ and $T=\vert t\vert^2$.

\begin{figure}
\centering\includegraphics[width=0.85\linewidth]{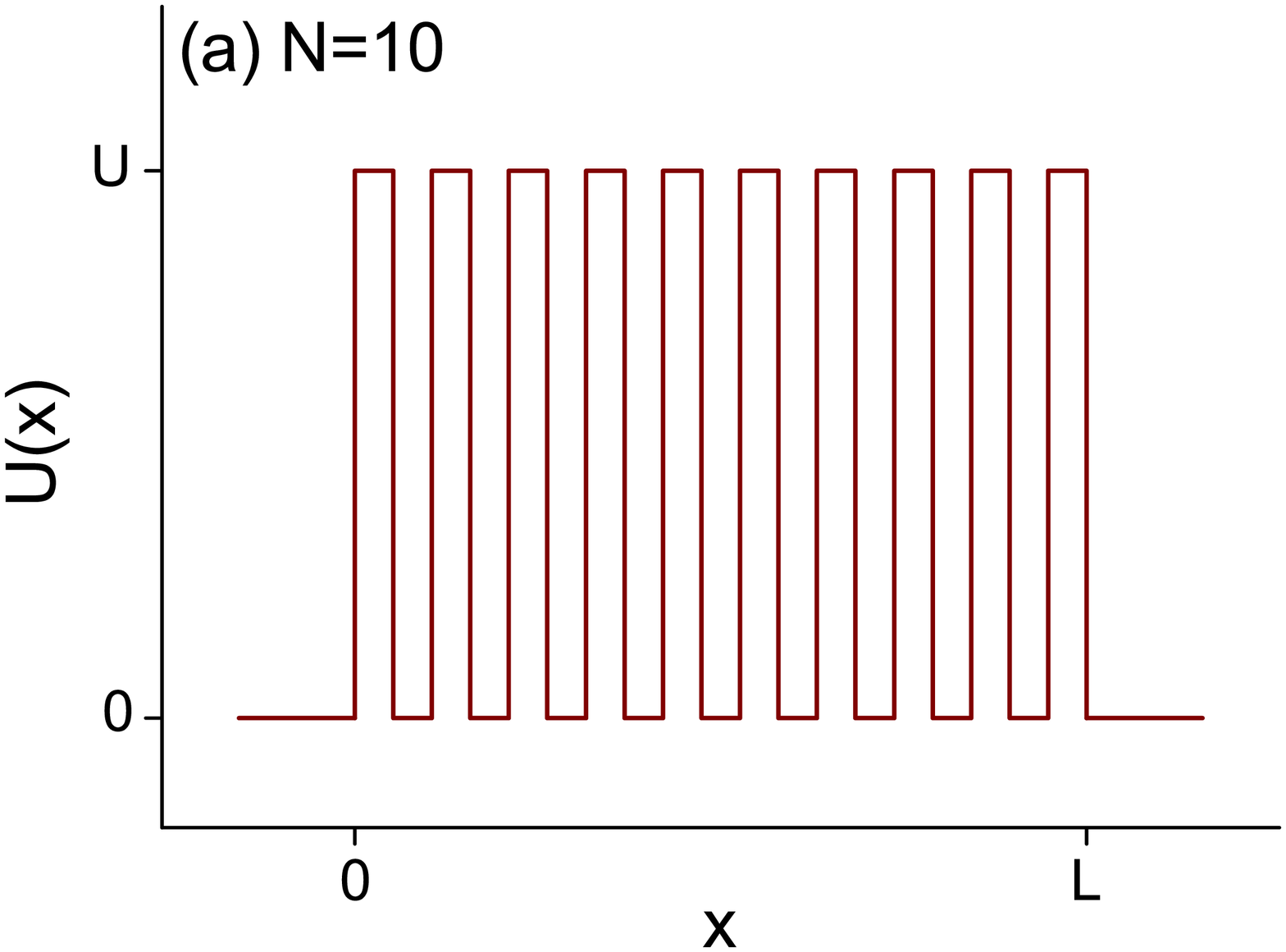}
\centering\includegraphics[width=0.85\linewidth]{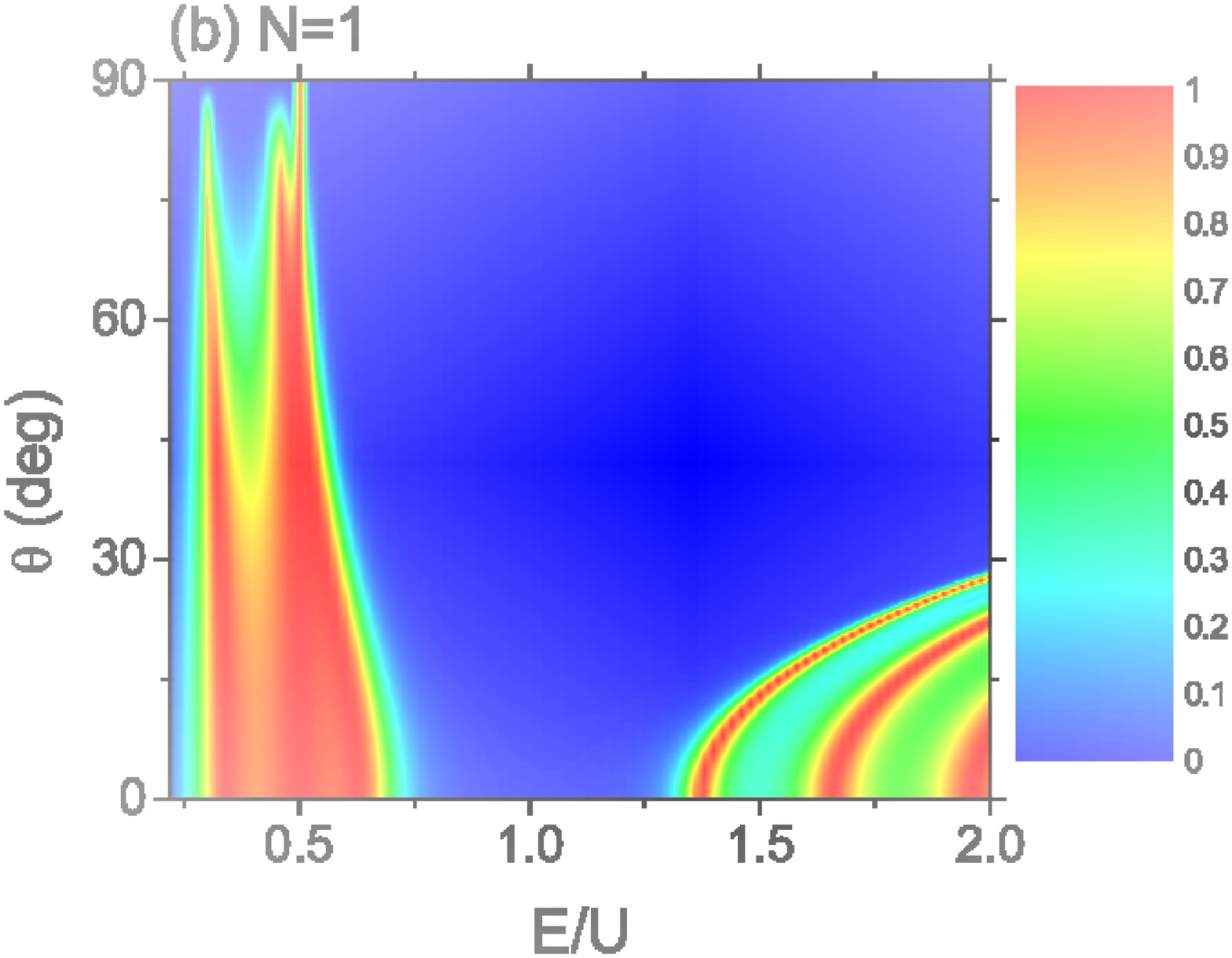}
\centering\includegraphics[width=0.85\linewidth]{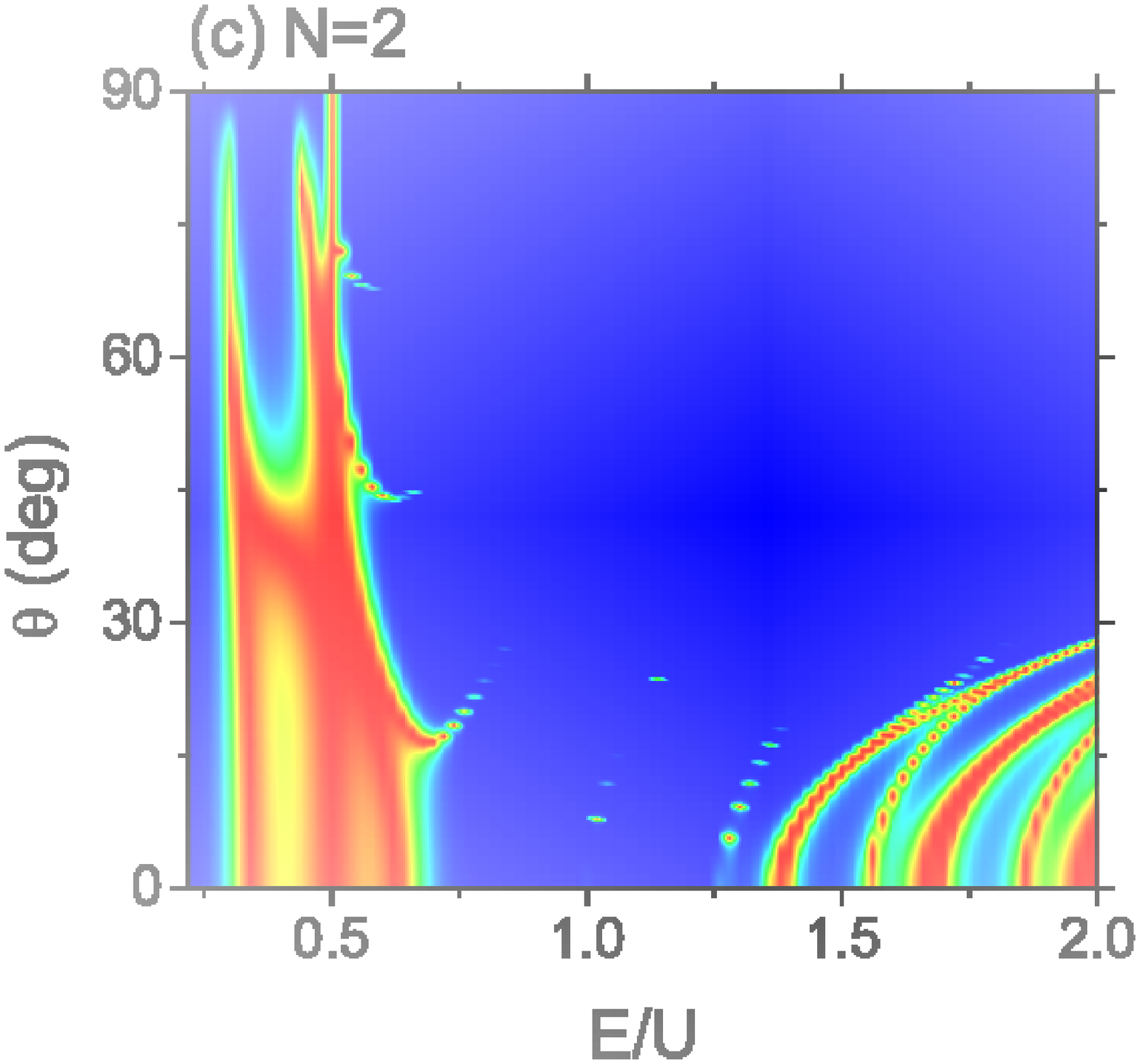}
\centering\includegraphics[width=0.85\linewidth]{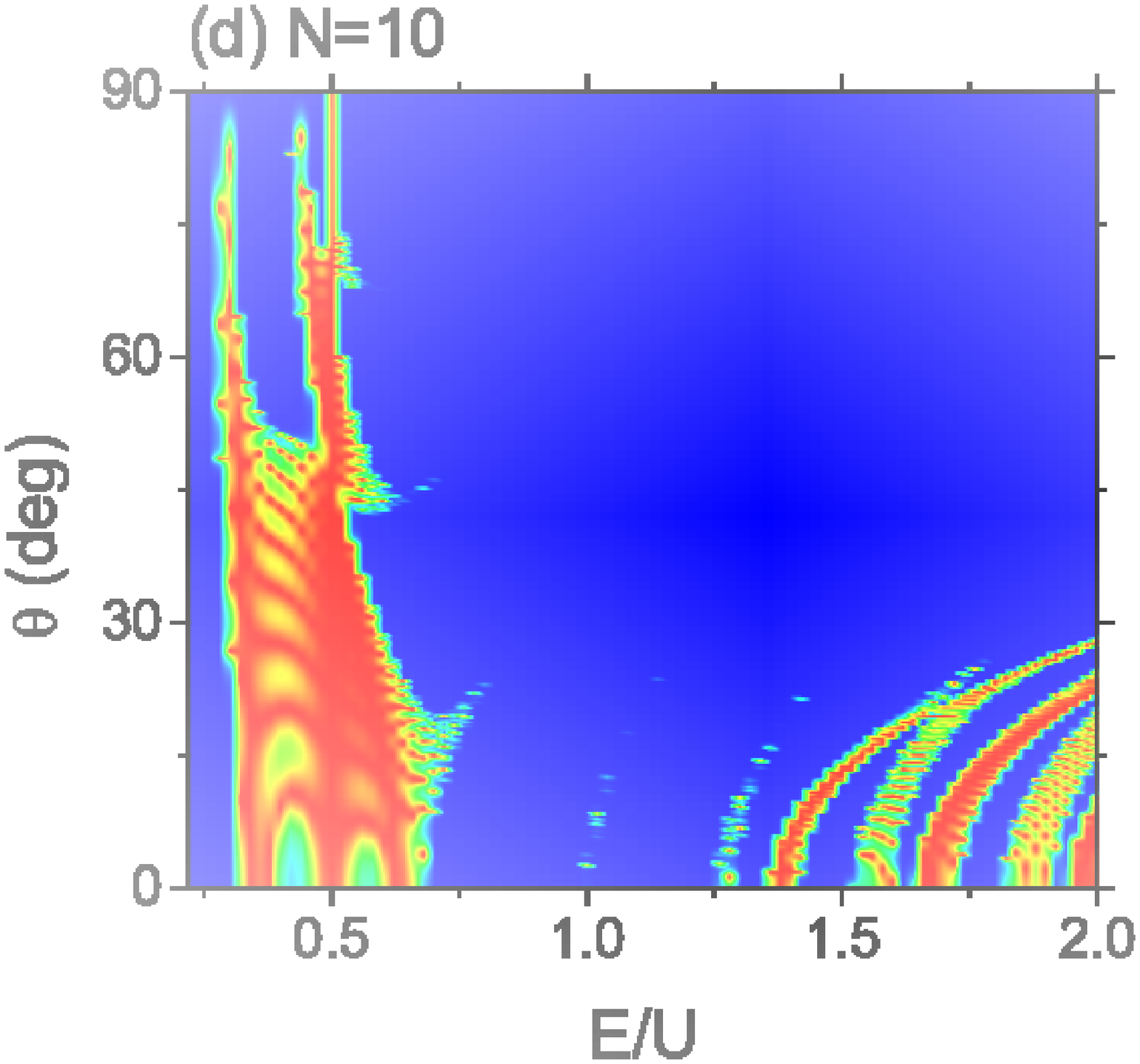}
 \caption{(a) Model structure of a periodic array of 10 identical potential barriers corresponding to (d).
 (b-d) Contour plots of the transmittance $T$ as a function of the incident angle $\theta$ and the particle energy
 normalized by the potential, $E/U$, for arrays of $N$ ($=1$, 2, 10) identical potential barriers when $mc^2/U=0.2$.
 The width of a single barrier, $w$, is the same as that
 of the free region between two neighboring barriers and satisfies $w=10\hbar c/U$.
 The transmittance is equal to 1 at $E/U=0.5$ for all $\theta$ and $N$. }
 \label{fig1}
 \end{figure}

\section{Numerical results}

The super-Klein tunneling occurs when $E=U/2$ regardless of the number of scattering interfaces. Therefore it will occur
for any arbitrary array of rectangular potential barriers of the same height.
In Fig.~\ref{fig1}(a), we illustrate the model structure of a periodic array of 10 identical potential barriers.
In Figs.~\ref{fig1}(b), \ref{fig1}(c) and \ref{fig1}(d), we show
the contour plots of the transmittance $T$ as a function of the incident angle and the particle energy
 normalized by the potential, $E/U$, for arrays of $N$ ($=1$, 2, 10) identical potential barriers when $mc^2/U=0.2$.
 The width of a single barrier, $w$, is the same as that
 of the free region between two neighboring barriers and satisfies $w=10\hbar c/U$.
We clearly observe that the transmittance is identically equal to 1 at $E/U=0.5$ for all $\theta$ and $N$.

\begin{figure}
\centering\includegraphics[width=0.85\linewidth]{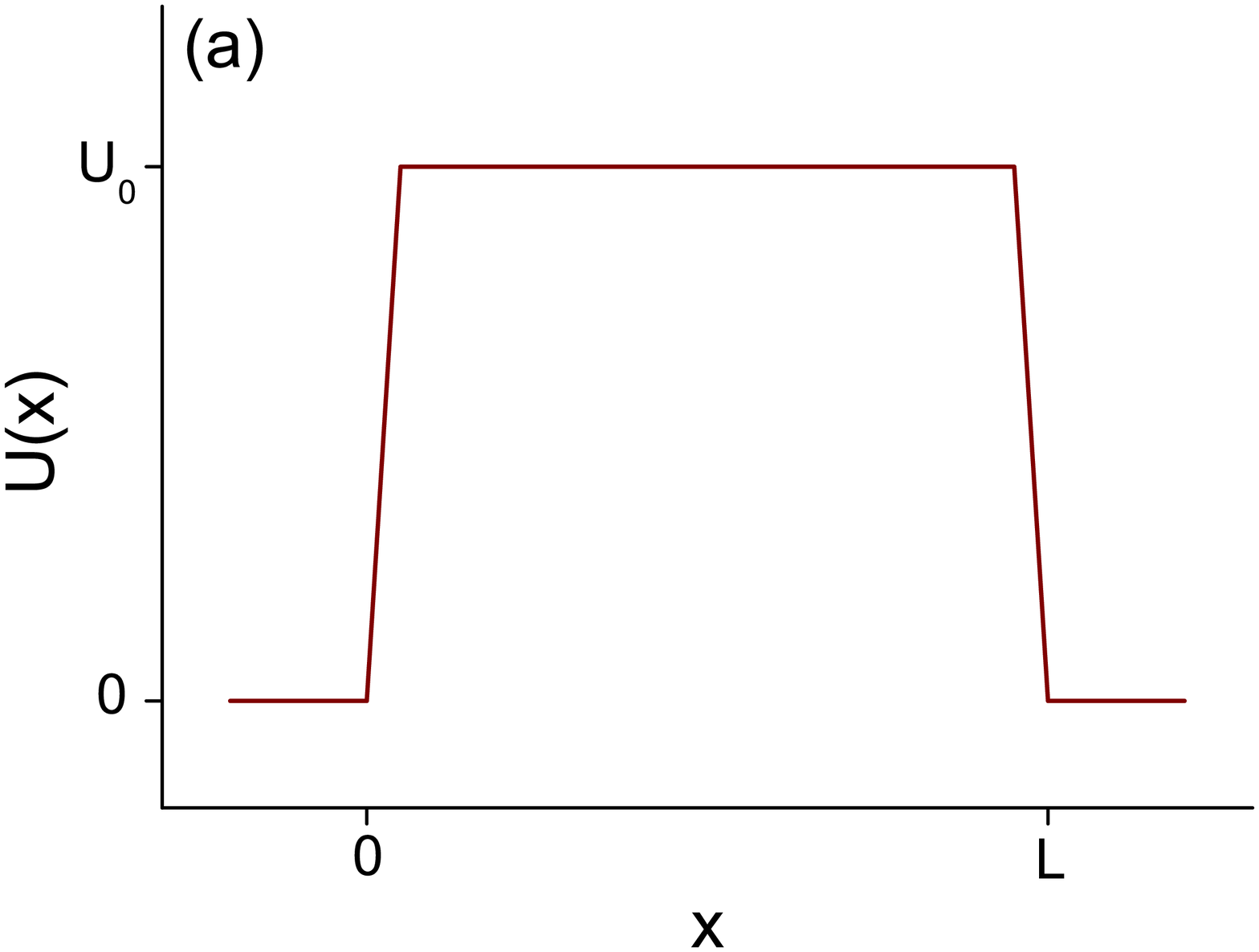}
\centering\includegraphics[width=0.85\linewidth]{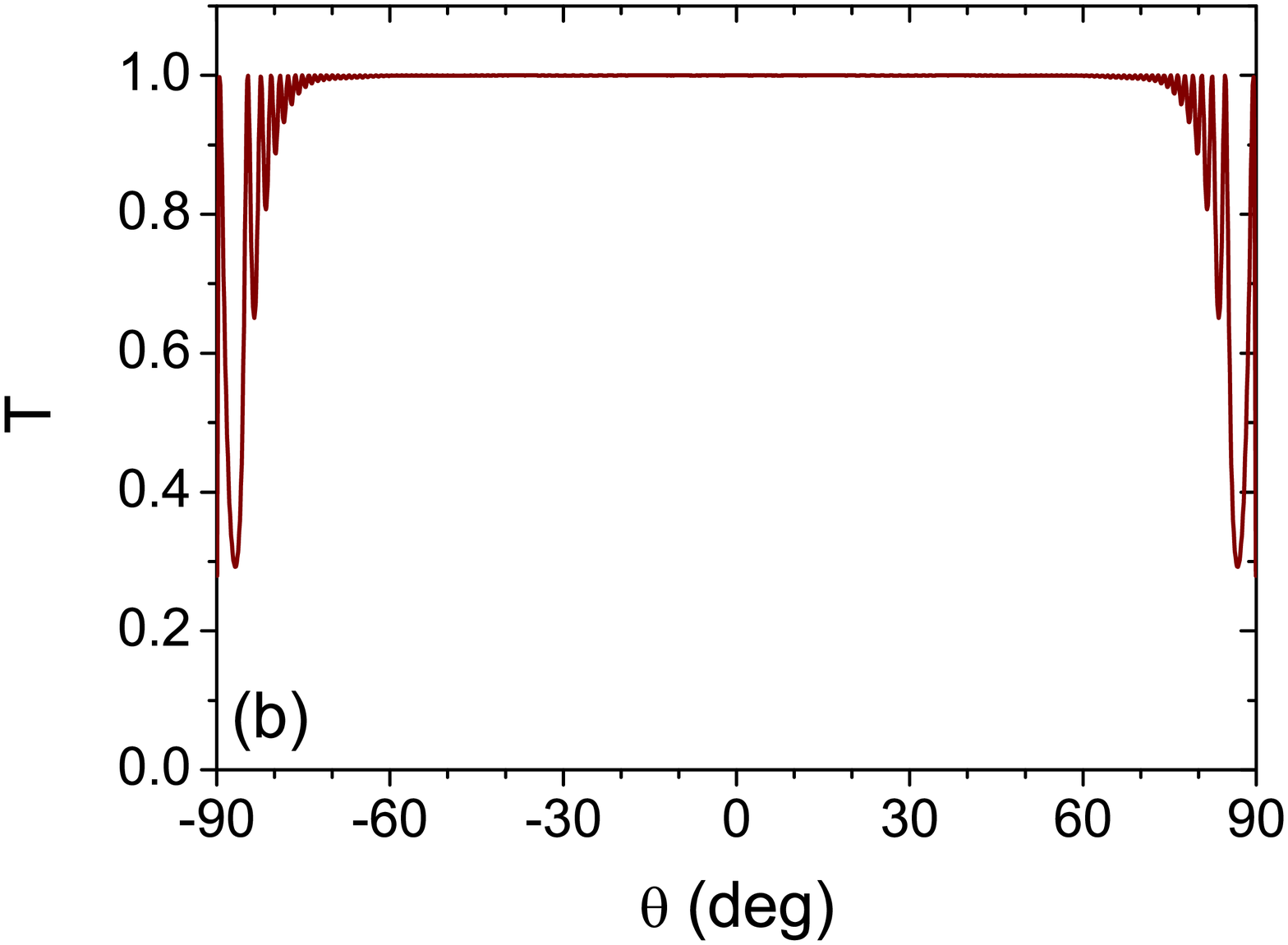}
 \caption{(a) Spatial configuration of the barrier $U$ of width $L$, which is equal to $U_0$ when $0.05\le x/L\le 0.95$
 and is given by $(U_0/0.05)x/L$ and $(U_0/0.05)(1-x/L)$ when $0\le x/L<0.05$ and
 $0.95<x/L\le 1$, respectively. (b) Transmittance versus incident angle for the barrier in (a).
 The particle energy and mass are given by $E/U_0=0.5$ and $mc^2/U_0=0.15$
 and the barrier width is given by $kL=300$.}
 \label{fig2}
 \end{figure}

In order to have perfect super-Klein tunneling, the scalar potential has to change from 0 to $U$ ($=2E$) discontinuously
at the interfaces.
However, this is obviously unphysical and we expect that there should exist a thin transition region where the potential
varies from 0 to $U$ in a continuous manner. Then, for a slab of finite thickness, a finite amount of reflection occurs
in the transition regions at the two interfaces and the Fabry-Perot-type interference of reflected waves can destroy the omnidirectional nature of the super-Klein tunneling.
In order to test this effect, we have calculated the transmittance for the slab of finite thickness with thin transition layers at the interfaces
depicted in Fig.~\ref{fig2}(a). This barrier of width $L$ is equal to $U_0$ when $0.05\le x/L\le 0.95$ and is given by $(U_0/0.05)x/L$ and $(U_0/0.05)(1-x/L)$ when $0\le x/L<0.05$ and
 $0.95<x/L\le 1$, respectively.
In Fig.~\ref{fig2}(b), we plot the transmittance versus incident angle when the particle energy and mass are given by $E/U_0=0.5$ and $mc^2/U_0=0.15$
 and the barrier width is given by $kL=300$. In the presence of reasonably thin transition layers, we find that perfect transmission
 is largely maintained, but for the case of grazing incidence with $\theta\to \pm 90^\circ$, there appears an oscillatory dependence on $\theta$
 due to Fabry-Perot interference
 and the transmission is not perfect.

\begin{figure}
\centering\includegraphics[width=0.85\linewidth]{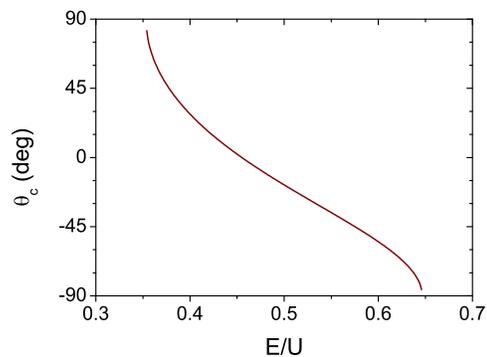}
 \caption{$\theta_c$ obtained from Eq.~(\ref{eq:ang}) plotted versus $E/U$ when $mc^2/U=0.1$ and $\tilde a=eA_y/U=0.3$.}
 \label{fig3}
 \end{figure}

\begin{figure}
\centering\includegraphics[width=0.85\linewidth]{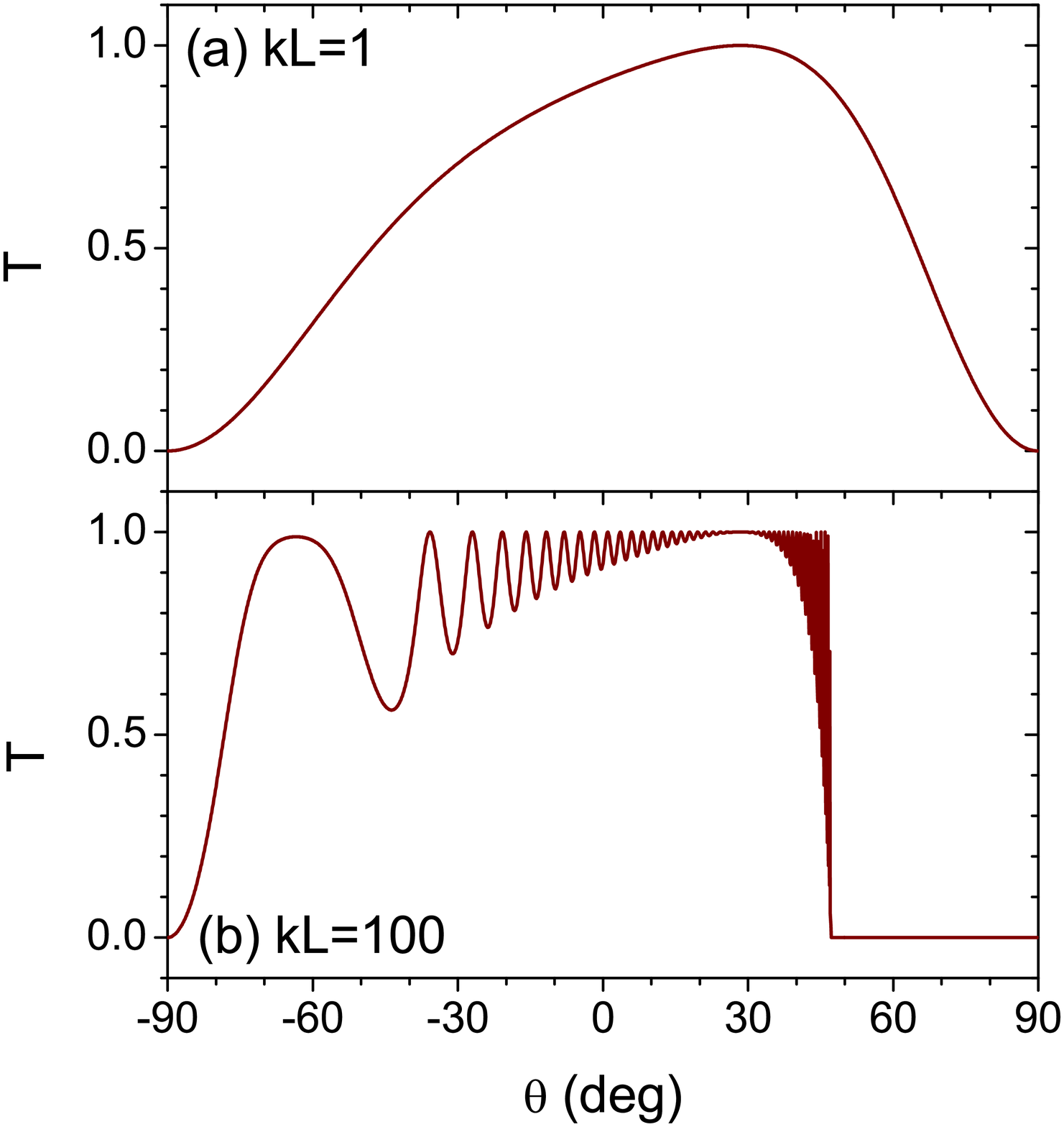}
 \caption{Transmittance versus incident angle for a barrier with both scalar and vector potentials, $U$ and $a$, of width $L$,
 when $U/E=2.5$, $a=0.9$, $mc^2/E=0.5$
 and (a) $kL=1$, (b) $kL=100$. The transmittance is equal to 1 at $\theta_c\approx 28.42^\circ$ in both cases.}
 \label{fig4}
 \end{figure}

We now consider the case where there are both scalar and vector potentials.
The angle $\theta_c$ given by Eq.~(\ref{eq:ang}) can be tuned continuously by tuning the value of $E$, $U$ or $a$.
We illustrate this in Fig.~\ref{fig3} by plotting $\theta_c$ versus $E/U$ when $mc^2/U=0.1$ and ${\tilde a}\equiv eA_y/U
=a\sqrt{(E/U)^2-(mc^2/U)^2}=0.3$. We notice that there exists a value of $E/U$ at which the angle $\theta_c$ vanishes.

Finally, in Fig.~\ref{fig4}, we plot the transmittance versus incident angle
for a barrier with both scalar and vector potentials, $U$ and $a$, of width $L$,
 when $U/E=2.5$, $a=0.9$, $mc^2/E=0.5$
 and $kL=1$, 100. In this situation, perfect transmission can occur at the incident angle satisfying
 Eq.~(\ref{eq:ang}), $\theta_c\approx 28.42^\circ$, which is precisely shown in Figs.~\ref{fig4}(a) and \ref{fig4}(b).
 When $L$ is large, there appear a large number of incident angles corresponding to Fabry-Perot resonances at which the transmission is perfect,
 in addition to $\theta_c$. In Fig.~\ref{fig4}(b) where $L$ is very small, these resonances are suppressed.

\section{Conclusion}
In this paper, we have studied the total transmission of Klein-Gordon particles
through a potential barrier based on the classical wave propagation theory.
We have proved that the omnidirectional total transmission phenomenon termed super-Klein tunneling
occurs in these systems
and derived a condition for total transmission of Klein-Gordon particles in the presence of both scalar and vector potentials.
The Klein-Gordon equation appears in a variety of wave propagation phenomena including
Alfv\'en waves in plasmas and acoustical waves in narrow tubes \cite{mf,forb,peet,ony}. It will be very interesting to design suitable experiments
in such systems to test our results.

\acknowledgments
This work has been supported by the National Research Foundation of Korea Grant (NRF-2018R1D1A1B07042629) funded by the Korean Government.


\end{document}